\documentclass[journal]{IEEEtran}
\usepackage{amsmath,amsfonts}
\usepackage{algorithmic}
\usepackage{array} 
\usepackage{textcomp}
\usepackage{stfloats}
\usepackage{url}
\usepackage{verbatim}
\usepackage{graphicx}
\def\BibTeX{{\rm B\kern-.05em{\sc i\kern-.025em b}\kern-.08em
    T\kern-.1667em\lower.7ex\hbox{E}\kern-.125emX}}
\usepackage{balance}

\usepackage{booktabs} 
\usepackage{tabularx}
\usepackage{threeparttable}
\usepackage{caption}
\usepackage{array}

\newcolumntype{C}{>{\centering\arraybackslash}X}
\newcolumntype{L}{>{\raggedleft\arraybackslash}X}
\newcolumntype{R}{>{\raggedright\arraybackslash}X}

\newcommand{\secref}[1]{Section~\ref{#1}}
\newcommand{\figref}[1]{Fig.~\ref{#1}}
\newcommand{\tabref}[1]{Table~\ref{#1}}
\newcommand{\algref}[1]{Algorithm~\ref{#1}}
\newcommand{\apdref}[1]{Appendix~\ref{#1}}

\usepackage{subfigure}
\usepackage{subcaption}
\usepackage[ruled, linesnumbered, vlined]{algorithm2e}  
\usepackage[colorlinks,linkcolor=red,anchorcolor=blue,citecolor=green]{hyperref}
\begin{document}
%\title{STI-SNN: High Energy-Efficiency Single-Timestep Inference \\ SNN Accelerator with Algorithm and Hardware Co-Design}
\title{STI-SNN: A 0.14 GOPS/W/PE Single-Timestep Inference \\ FPGA-based SNN Accelerator with \\ Algorithm and Hardware Co-Design}
\author{Kainan Wang, Chengyi Yang, Chengting Yu, \\Yee Sin Ang, Bo Wang,~\IEEEmembership{Senior Member,~IEEE} and Aili Wang,~\IEEEmembership{Member,~IEEE} 
\thanks{
This work was supported in part by National Natural Science Foundation of China under grant 62304203, in part by Natural Science Foundation of Zhejiang Province, China under Grant LQ22F010011, and in part by ZJU-UIUC Dynamic Research Enterprise for HeterogeneouslY integrated BRain InspireD computing (HYBRID). \textit{(Kainan Wang and Chengyi Yang equally contributed to this work, corresponding author: Aili Wang.)} 
}
\thanks{Kainan Wang, Chengyi Yang, Chengting Yu, and Aili Wang are with the Zhejiang University-University of Illinois Urbana-Champaign Institute, Zhejiang University, Haining, China. (e-mail: \{kainan.22, chengyi.22, chengting.21, ailiwang\}@intl.zju.edu.cn).}

\thanks{Yee Sin Ang and Bo Wang are with the Information Systems Technology and Design,
Singapore University of Technology and Design, Singapore (e-mail: \{yeesin\_ang, bo\_wang\}@sutd.edu.sg).}
}
% the College of Information Science and Electronic Engineering 
\maketitle

\begin{abstract}
Brain-inspired Spiking Neural Networks (SNNs) have attracted attention for their event-driven characteristics and high energy efficiency. However, the temporal dependency and irregularity of spikes present significant challenges for hardware parallel processing and data reuse, leading to some existing accelerators falling short in processing latency and energy efficiency.
To overcome these challenges, we introduce the STI-SNN accelerator, designed for resource-constrained applications with high energy efficiency, flexibility, and low latency. The accelerator is designed through algorithm and hardware co-design.
Firstly, STI-SNN can perform inference in a single timestep. At the algorithm level, we introduce a temporal pruning approach based on the temporal efficient training (TET) loss function. This approach alleviates spike disappearance during timestep reduction, maintains inference accuracy, and expands TET's application. In hardware design, we analyze data access patterns and adopt the output stationary (OS) dataflow, eliminating the need to store membrane potentials and access memory operations. Furthermore, based on the OS dataflow, we propose a compressed and sorted representation of spikes, then cached in the line buffer to reduce the memory access cost and improve reuse efficiency.
Secondly, STI-SNN supports different convolution methods. By adjusting the computation mode of processing elements (PEs) and parameterizing the computation array, STI-SNN can accommodate lightweight models based on depthwise separable convolutions (DSCs), further enhancing hardware flexibility.
Lastly, STI-SNN also supports both inter-layer and intra-layer parallel processing. For inter-layer parallelism, we adopt a streaming architecture and introduce a layer-wise pipelining approach to reduce overall latency. We also develop a latency model for the convolution layer and propose an intra-layer parallelism strategy for output channels, further reducing the inference latency.
% results
Algorithmic experiments demonstrate that the single timestep inference accuracy of STI-SNN is competitive. On the CIFAR10 dataset, the accuracies of the spiking ResNet19 and VGG16 reach 93.74\% and 93.76\%, respectively. 
We implement three models (SCNN3, SCNN5 and vMobileNet) on the Xilinx Zynq Ultrascale ZCU102 FPGA platform. The hardware implementation results indicate that, within a single timestep inference, SCNN5 can save 126KB of on-chip storage resources, and compared to two timesteps, the energy consumption to infer the same samples is approximately halved. For SCNN3 and SCNN5, the accelerator's inference speed increased by 3.91$\times$ and 4$\times$, and the efficiency improved by 3.64$\times$ and 3.49$\times$, with single PE efficiencies reaching 0.19 and 0.14 GOPS/W/PE, respectively.
\end{abstract}
\begin{IEEEkeywords}
Algorithm and Hardware Co-Design, FPGA, Low Latency and High Energy Efficiency, Spiking Neural Networks, Single-Timestep Inference.
\end{IEEEkeywords}

\section{Introduction}  
\IEEEPARstart{R}{ecently}, Spiking Neural Networks (SNNs) have rapidly developed as a promising alternative to Artificial Neural Networks (ANNs), thanks to their ultra-low power consumption. Unlike deep neural networks (DNNs), which require extensive computational and storage resources, SNNs transmit information through spikes among neurons. This discrete spiking behavior not only introduces event-driven characteristics (i.e., the membrane potential of neurons is updated only when input spikes are received) but also achieves high energy efficiency by replacing multiply-accumulate (MAC) operations with addition.

Although SNN inference is more energy efficient than ANN, the von Neumann architecture of CPUs/GPUs does not fully support the temporal dependency and event-driven characteristics of SNNs. Therefore, many researchers have developed high-throughput and energy-efficient SNN processors. Examples include IBM's TrueNorth~\cite{a1}, Intel's Loihi~\cite{a2}, Xilinx's S2N2~\cite{a3}, Neurogrid~\cite{a4}, and Tianjic~\cite{a5}. The tight coupling of computation and architecture in SNNs requires an algorithm and hardware co-design. 
This paper aims to achieve a low-latency and high-energy-efficiency SNN accelerator while maintaining inference accuracy through an algorithm and hardware co-design for applications with limited hardware resources.

The computation of SNNs requires multiple timesteps, and as the number of timesteps increases, the computational cost of neurons and the memory access cost of membrane potentials also gradually increase. This reduces the inference efficiency and poses significant challenges for hardware deployment. Recent studies have employed various temporal pruning techniques to optimize the energy consumption of inference while achieving acceptable accuracy. However, we find that most state-of-the-art (SOTA) approaches require hardware support for complex computation units~\cite{b1,b6,b7,b8}, significantly increase training costs~\cite{b3, b4, b5}, and some are even inadequate for handling complex tasks~\cite{b2}. 
For instance, Xie et al.~\cite{b1} propose a layer-wise configurable timesteps (LCTs) approach that addresses temporal redundancy by analyzing the Shannon bits of each layer. However, the computation of principal component analysis (PCA) incurs additional overhead for the hardware. 
Lew et al.~\cite{b2} introduce temporal neuron pruning based on membrane voltage. Although this approach incurs relatively low hardware overhead, its effectiveness in optimizing the timesteps is quite limited, as it primarily addresses the computation of less important neurons by skipping them. 
The authors of ~\cite{b3,b4,b5} perform temporal pruning operations, and to address spike disappearance caused by reduced timesteps, they incorporate neuron parameters (such as membrane potential thresholds and leak parameters) into the training process, which increases training costs. Additionally, the approach~\cite{b3} is limited by the local learning algorithm Spike-Timing-Dependent Plasticity (STDP) and its simple network structure, making it challenging to handle more complex tasks. Chowdhury et al.~\cite{b5} achieve SNNs with a single inference timestep, but the proposed iterative initialization and retraining (IIR) significantly increased training costs. 
Compared to these fixed timesteps compression techniques, there are some alternative approaches~\cite{b6,b7,b8} that can support dynamic timesteps by calculating metrics such as maximum entropy or confidence. Although these approaches can achieve a better balance between accuracy and timesteps, they are not suitable for circuit designs with streaming architectures. Furthermore, spike disappearance limits schemes with fewer timesteps~\cite{b9, b10}, and these approaches introduce complex calculations (such as exponentials, logarithms, and division), posing significant challenges for hardware deployment.

In hardware design, the most advanced SNN hardware designs aim to fully exploit sparse spikes. However, the temporal dependency and irregularity of these spikes require neurons to be updated at each timestep~\cite{c1,c2}. This sequential processing increases both the computational burden and the frequency of access to membrane potential memory, while also affecting the effective implementation of parallel processing and data reuse~\cite{a1,a2}. As a result, issues like lengthy inference delays and low energy efficiency persist. Lee et al.~\cite{c5} and Chen et al.~\cite{c6} have explored spatiotemporal parallelism by unrolling computations in both temporal and spatial dimensions, achieving significant acceleration.  However, parallelization across multiple timesteps violates the temporal dependency sequential nature of neuron membrane potential updates. Liu et al.~\cite{c7} introduced a temporal-parallel dataflow and a bucket-sort based dispatcher, enhancing the performance of the SATO accelerator. Narayanan et al.~\cite{c8} introduced a compressed, time-stamped, and sorted spike representation, along with a tailored dataflow that enhances accelerator performance. Nevertheless, the temporal encoding methods adopted in ~\cite{c7, c8} may compromise accuracy. The aforementioned methods can achieve high throughput through parallel processing. However, there remains the challenge of low energy efficiency caused by the frequent movement of membrane potential. Therefore, it is crucial to manage the reuse of membrane potential in SNNs. Chen et al.~\cite{c9} proposed a row stationary (RS) dataflow for their Eyeriss accelerator, which can reduce the energy consumption of moving weights and input data. However, the partial sums are not fully accumulated before being offloaded to the global buffer, which increases the energy consumption associated with the movement of the neuron membrane potential at each timestep.

\begin{figure*}[tbp]
    \centering
    \includegraphics[width=\linewidth]{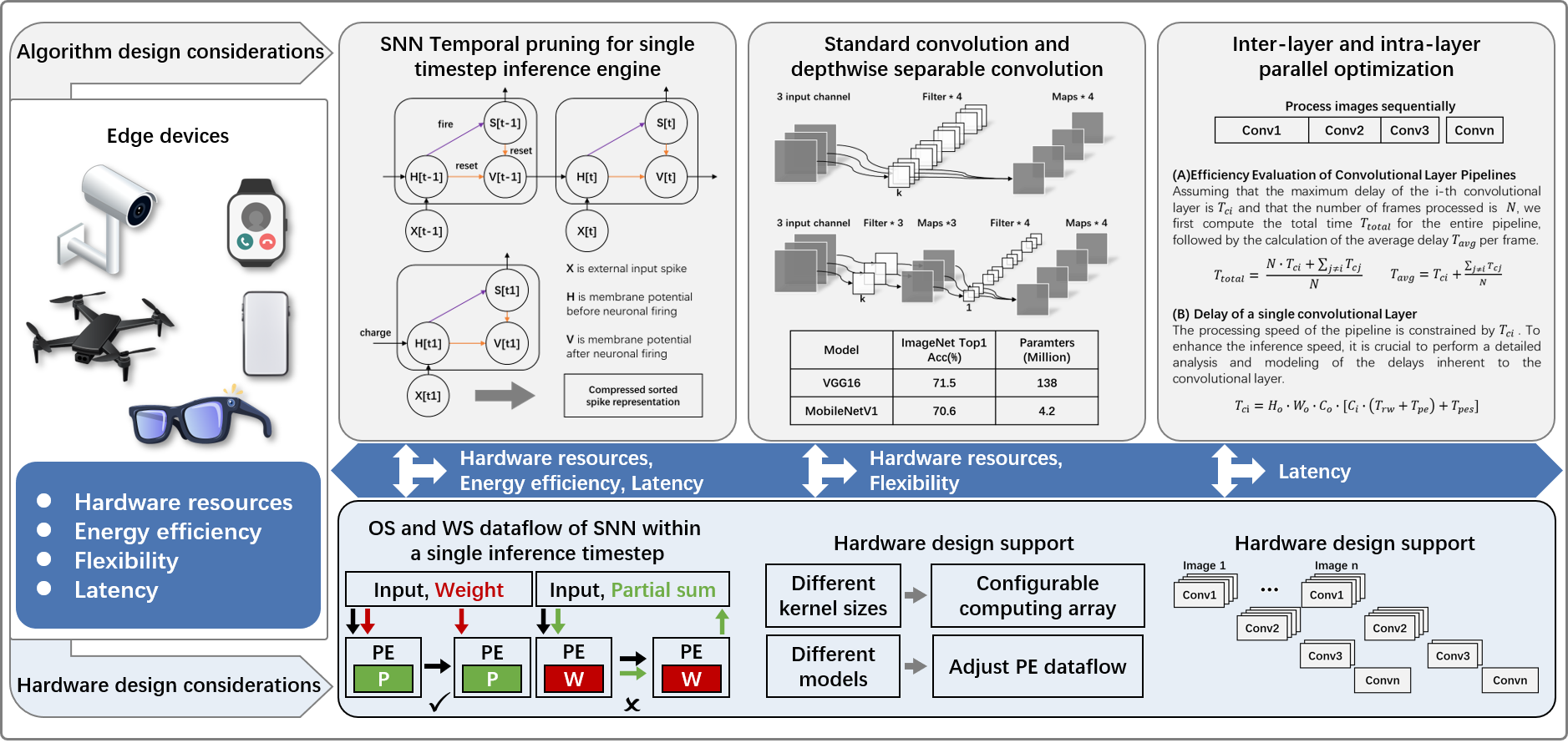}
    \caption{
    Algorithm and hardware co-design based STI-SNN accelerator.
    }
    \label{fg:ah}
\end{figure*}

To achieve low-latency, high-accuracy inference in resource-constrained environments while ensuring high energy efficiency and hardware design flexibility, we introduce STI-SNN, a solution that leverages algorithm and hardware co-design, as shown in \figref{fg:ah}.
\textbf{Firstly}, STI-SNN supports inference with a single timestep. At the algorithm level, we introduce a novel temporal pruning approach based on the temporal efficient training (TET) loss function~\cite{c10}, which maintains inference accuracy within a single timestep while avoiding the introduction of complex computational units, thus simplifying hardware requirements. In hardware design, we adopt output stationary (OS) dataflow rather than weight stationary (WS) dataflow when designing standard convolutional networks. By jointly considering the design of the algorithm and hardware, we can eliminate the need for external memory storage of membrane potentials and reduce related memory access operations. Furthermore, based on OS dataflow, we propose a compressed and sorted representation of input spikes, which are then cached into the line buffer, reducing memory access costs and improving on-chip data reuse efficiency.
\textbf{Secondly}, STI-SNN supports different convolution methods for SNN models. Many lightweight models have already been designed and explored at the algorithm level. For example, models based on depthwise separable convolutions (DSCs) are more compact, achieving similar accuracy with fewer parameters. To enable hardware compatibility with different convolution methods, we adjust the computation mode of processing elements (PEs) and parameterize the configuration of the computation array, thereby improving hardware flexibility and adapting to the needs of various algorithmic models.
\textbf{Lastly}, the hardware design of STI-SNN supports both inter-layer and intra-layer parallel data processing. For inter-layer parallelism, we adopt a streaming architecture and implement a layer-wise pipelining scheme to reduce overall latency. By modeling and analyzing the latency of the convolutional layers, we identify that the inference latency bottleneck occurs in the convolution layer with the highest delay. To address this, we propose an intra-layer parallelism optimization strategy for output channels, further reducing inference latency and improving overall efficiency in the hardware implementation.
Through the above algorithm and hardware co-design, STI-SNN achieves high inference accuracy while effectively reducing latency and energy consumption, providing new insights and practical solutions for designing efficient and flexible SNN accelerators.

The algorithmic experimental results show that with single timestep inference, the accuracy of the CIFAR10 dataset on ResNet19 and VGG16 reaches 93.74\% and 93.76\%, respectively. Then, we evaluate the STI-SNN accelerator on FPGA, and the results indicate that SCNN5 saves 126KB of on-chip storage resources during single timestep inference. For SCNN3 and SCNN5, the accelerator's inference speed increased by 3.91$\times$ and 4$\times$, and efficiency improved by 3.64$\times$ and 3.49$\times$, with single PE's efficiency reaching 0.19 and 0.14 GOPS/W/PE, respectively.

The remainder of the paper is organized as follows. \secref{background} introduces the background of SNNs, including neurons and learning rules, and then compares and analyzes the memory access costs of data in different dataflows. \secref{algorithm} presents the single timestep inference approach designed for the compute engine, along with the temporal pruning experimental results. \secref{hardware} first provides an overview of the overall hardware architecture and the OS dataflow of the standard convolutional layer, followed by a discussion of the technologies corresponding to the goals of high energy efficiency, flexibility, and low latency. \secref{results} analyzes and compares the hardware experimental results. Finally, \secref{conclusion} concludes the paper.

\section{Background} \label{background}
\subsection{SNN Neurons}
This paper adopts a simplified model of neuron dynamics, where the leaky integrate and fire (LIF)~\cite{lif} model has been widely implemented in SNNs. The membrane potential of LIF neurons leaks over time and accumulates input spikes at each timestep. When the membrane potential exceeds a certain threshold, the neuron fires a spike. The dynamic process of the membrane potential $u$ of the LIF neuron over time $t$ can be described by \eqref{EQ_LIF}, 
\begin{equation}
\label{EQ_LIF}
\tau_{m} \frac{du}{dt}=-u+R I(t), \quad u<V_{th}.
\end{equation} 
where $V_{th}$ represents the threshold, $I$ denotes the input current, $R$ is the resistance, and $\tau_{m}$ represents the time constant of the membrane potential. When the membrane potential $u$ exceeds the threshold $V_{th}$, a spike is generated, after which the membrane potential is reset to the reset potential $u_{r}$, which is set to 0 in this paper. We adopt the discrete form of \eqref{EQ_LIF}, which is divided into three stages.

\subsubsection{Input Current Accumulation Phase}
At each discrete timestep, all presynaptic spikes are accumulated, as described in \eqref{EQ_LIF_sub1}, $w_{ij}$ represents the synaptic weight from neuron $j$ to neuron $i$, and $b_i$ is the bias. 
\begin{equation}
\label{EQ_LIF_sub1}
I[t]=\sum_{j} w_{i j} s_{j}[t]+b_{i}.
\end{equation} 
\subsubsection{Membrane Potential Update Phase}
The neuron's membrane potential is updated at each timestep based on the input current and the historical membrane potential, as described in \eqref{EQ_LIF_sub2}.
\begin{equation}
\label{EQ_LIF_sub2}
u_{i}[t]=\left(1-\frac{1}{\tau_{m}}\right) u_{i}[t-1]+I[t].
\end{equation} 
\subsubsection{Output Spike Generation Phase}
When the membrane potential reaches the threshold, the neuron fires an output spike and resets the membrane potential, as described in \eqref{EQ_LIF_sub3}.
\begin{equation}
\label{EQ_LIF_sub3}
\left(u_{i}[t+1], s_{i}[t+1]\right)=\left\{\begin{array}{l}
\left(u_{i}[t], 0\right), \quad u_{i}[t]<V_{t h} \\
(0,1), \quad u_{i}[t] \geq V_{t h}.
\end{array}\right. 
\end{equation} 
 
In these three stages, we have two key observations. Due to the high degree of synaptic connectivity and the large number of neurons, the input current accumulation phase dominates the total computational cost. The membrane potential update phase has stringent storage requirements, and the membrane potential needs to be read and written back at each timestep.

\subsection{SNN Learning Rules}
We train SNN using spatio-temporal backpropagation (STBP)~\cite{STBP}. During the backpropagation process, the non-differentiable activation term $da/du$ in \eqref{EQ_SDT1} is replaced by a surrogate gradient (SG)~\cite{SG}. Here, we adopt the default SG function in the SpikingJelly framework~\cite{spikingjelly}. Regarding the loss function, the most widely used is standard direct training (SDT), as shown in \eqref{EQ_SDT2}. Based on the chain rule, we can obtain the gradient of the weights, as shown in \eqref{EQ_SDT3}.

\begin{equation}
\label{EQ_SDT1}
\frac{\partial L}{\partial W}=\sum_{t} \frac{\partial L}{\partial \boldsymbol{a}(t)} \frac{\partial \boldsymbol{a}(t)}{\partial \boldsymbol{u}(t)} \frac{\partial \boldsymbol{u}(t)}{\partial \boldsymbol{I}(t)} \frac{\partial \boldsymbol{I}(t)}{\partial W}.
\end{equation} 
\begin{equation}
\label{EQ_SDT2}
\mathcal{L}_{\mathrm{SDT}}=\mathcal{L}_{\mathrm{CE}}\left(\frac{1}{T} \sum_{t=1}^{T} \boldsymbol{O}(t), \boldsymbol{y}\right).
\end{equation} 
\begin{equation}
\label{EQ_SDT3}
\frac{\partial \mathcal{L}_{\mathrm{SDT}}}{\partial W}=\frac{1}{T} \sum_{t=1}^{T}\left[S\left(\boldsymbol{O}_{\text {mean }}\right)-\hat{\boldsymbol{y}}\right] \frac{\partial \boldsymbol{O}(t)}{\partial W}.
\end{equation} 

\subsection{Data Reuse Efficiency Analysis} 
Taking standard convolution as an example, we analyze the memory access counts for input spikes, weights, and partial sums in a single convolution module under OS and WS dataflows. 
% input
In the OS dataflow, a single output pixel $P(W_o, H_o, C_o)$ requires $C_i \times K_w \times K_h$ memory accesses for the input. With a total of $C_o \times W_o\times H_o$ output pixels, the total memory accesses for the input is $C_i \times K_w \times K_h \times C_o \times W_o\times H_o \times T$. In the WS dataflow, processing the weights of a single channel requires $K_w \times K_h \times W_o \times H_o$ memory accesses for the input data. With a total of $C_i \times C_o$ weight channels, the total memory accesses for the input data is $K_w \times K_h \times W_o \times H_o \times C_i \times C_o \times T$.
% weights
Similarly, in the OS dataflow, a single output pixel requires $C_i \times K_w \times K_h$ memory accesses for weights. Since there are $C_o \times W_o \times H_o$ output pixels, the total memory accesses for weights is $C_i \times K_w \times K_h \times C_o \times W_o \times H_o \times T$. In the WS dataflow, each channel's weights need to be accessed only once, resulting in a total of $C_i \times K_w \times K_h \times C_o \times T$ memory accesses for the weights.
% partial sum
In the OS dataflow, when inferring at a single timestep, there is no need to store partial sums, resulting in a total of $C_o \times W_o \times H_o \times(T-1)$ memory accesses for partial sums. In the WS dataflow, $C_i \times C_o \times W_o \times H_o \times T$ memory accesses are required. \tabref{tab:al_t1} summarizes the above analysis.

As shown in \tabref{tab:al_t1}, the memory access cost for all data in both dataflows exhibits a linear relationship with the timesteps $T$. As $T$ increases, the data reuse efficiency decreases. Additionally, if a single memory access only includes one input spike or weight value, it will reduce bandwidth utilization and increase memory access overhead.
% compare ws and os
In comparing OS and WS, the number of memory accesses for weights in WS is $W_o \times H_o$ times greater than in OS, yet the required storage size for weights is the same in both dataflows. Moreover, we find that when inferring at a single timestep, OS is more suitable because it does not require storing and accessing partial sums, whereas WS still does.
Consequently, from the perspective of algorithm and hardware co-design, in resource-constrained environments, we aim to reduce the number of off-chip memory accesses, improve the efficiency of on-chip data reuse, and further enhance the system's throughput and energy efficiency. At the algorithm level, we reduce the inference latency to a single timestep, and at the hardware level, we select and optimize the OS dataflow.

\begin{table}[tbp]
\renewcommand{\arraystretch}{1.25}
\centering\caption{\label{tab:al_t1}Memory Access Counts for Input spikes, Weights, and Partial Sums under OS and WS Dataflows.}
\begin{tabularx}{\columnwidth}{CCC} 
\toprule
& OS & WS \\
\midrule
Input spikes & $C_i \times K_w \times K_h \times C_o \times W_o\times H_o \times T$ & $K_w \times K_h \times W_o \times H_o \times C_i \times C_o \times T$ \\
Weights & $C_i \times K_w \times K_h \times C_o \times W_o \times H_o \times T$ & $C_i \times K_w \times K_h \times C_o \times T$ \\
Partial Sums & $C_o \times W_o \times H_o \times(T-1)$ & $C_i \times C_o \times W_o \times H_o \times T$ \\ 
\bottomrule
\end{tabularx}\end{table}

\section{Proposed Algorithm for Single Timestep Inference Engine} \label{algorithm}
\subsection{SNN Inference Timesteps Compression} 
\subsubsection{Accuracy Trend in SDT} 
\begin{figure}[tbp]
    \centering
    \includegraphics[width=0.95\columnwidth]{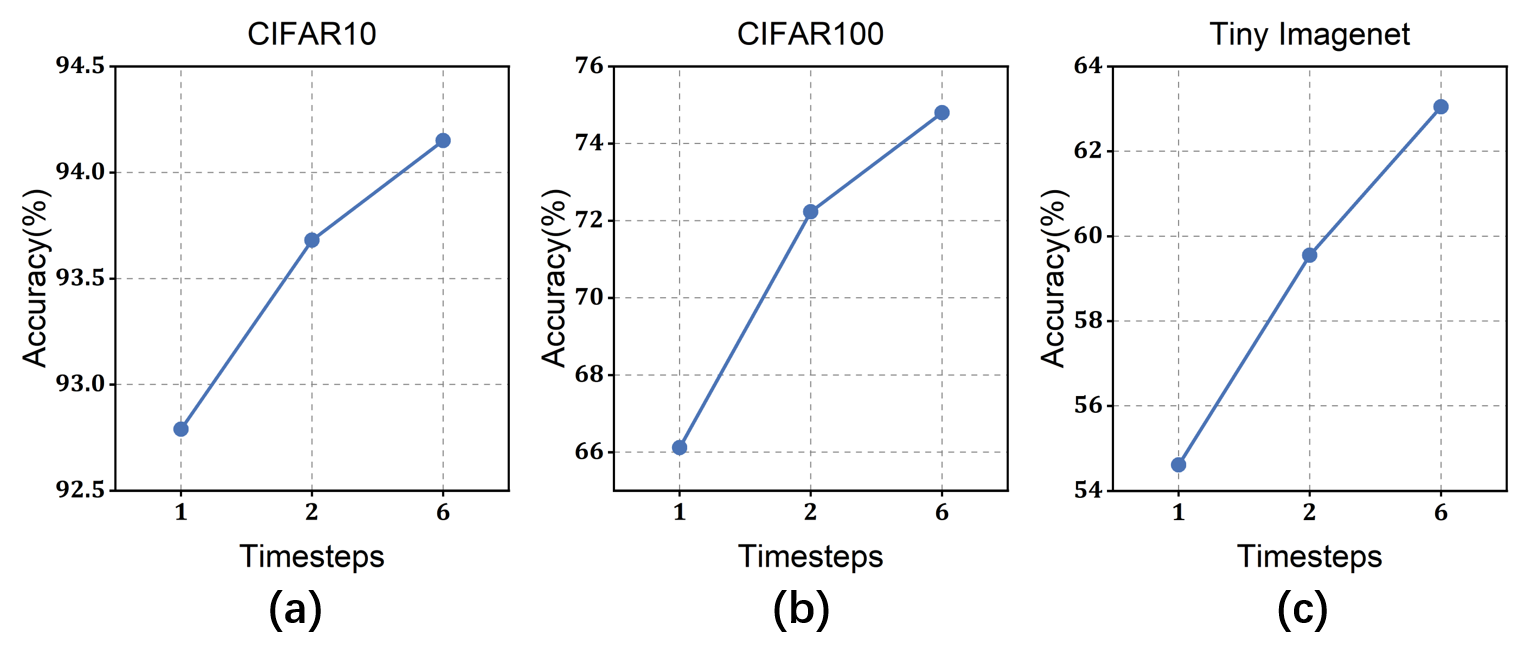}
    \caption{ 
    Impact of inference timesteps on the SNN models' accuracy: testing spiking VGG16 on CIFAR10 and CIFAR100, and ResNet34 on Tiny ImageNet.   
    }
    \label{fg:t-acc}
\end{figure}

The timesteps of SNNs lead to a trade-off between accuracy and efficiency. Generally, increasing the number of timesteps results in more spike events, which helps SNNs capture greater temporal information and improve task performance. However, this also results in higher energy consumption and reduced efficiency due to the increased spike computations and longer delays.
Here, we present several accuracy versus timesteps curves. \figref{fg:t-acc} shows that SDT with a single timestep results in significant accuracy loss on the CIFAR100 and Tiny ImageNet datasets, a phenomenon consistent with findings in studies~\cite{b6,b8}. This indicates that training an SNN model with only a single timestep is not feasible. Consequently, we next explore the feasibility of reducing the inference delay directly from the initial timesteps of 6 to 1 (training with 6 timesteps and inference with 1 timestep). 

\subsubsection{Directly Reduce Inference Timesteps}  
\begin{figure}[tbp]
    \centering
    \includegraphics[width=0.95\columnwidth]{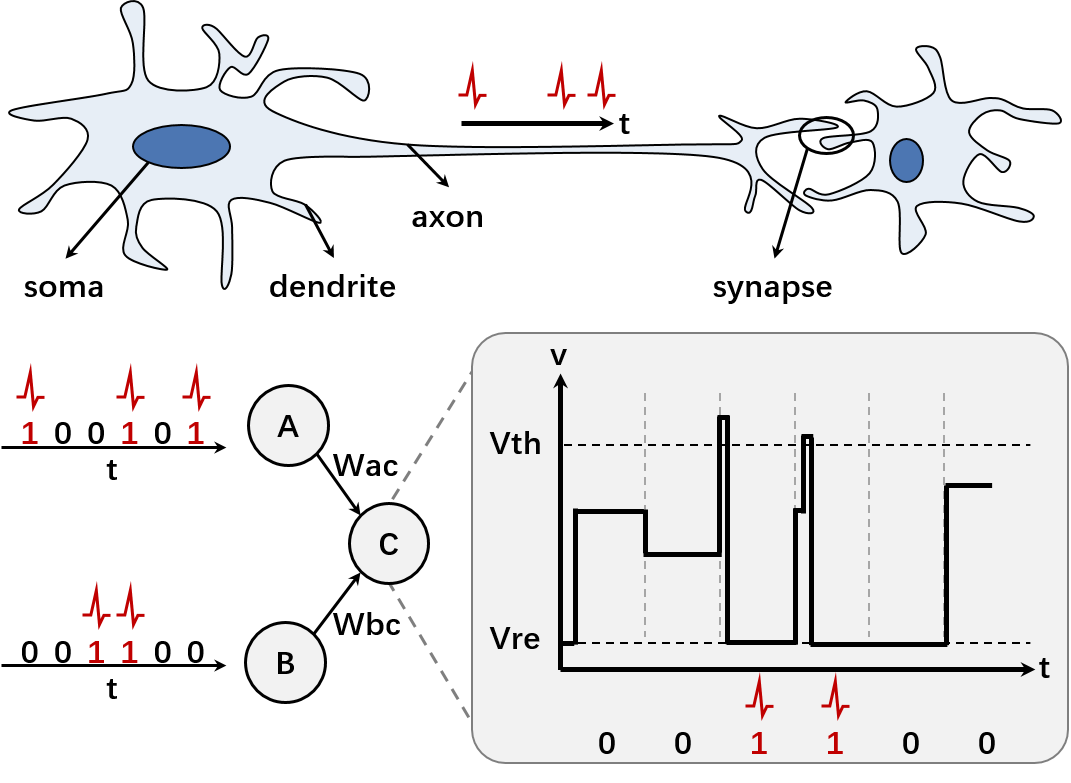}
    \caption{
    Impact of inference timesteps on the neuron activity.
    }
    \label{fg:3-neurons}
\end{figure}

As shown in \figref{fg:3-neurons}, neuron C receives inputs from neurons A and B in the preceding layer, with synaptic weights trained through SDT over 6 timesteps, it fires a spike when its membrane potential exceeds a defined threshold.
\figref{fg:3-neurons} also shows that neuron C is sensitive to changes in input timesteps, as directly reducing the inference timesteps can prevent it from firing, which leads to poor generalization. This could be because when the SDT approaches a local minimum, the error term in \eqref{EQ_SDT3} approximates 0 at all time points, and the gradient mismatch in SG \cite{missmatch} can result in an extremely small accumulated momentum, making it difficult for the model to escape the sharp local minimum. Consequently, this limitation on weight updates restricts the model's capacity to learn the complex features of the input data.
We draw inspiration from related research and adopt the TET method, aimed at enhancing the model's generalization and improving its adaptability to variations in input timesteps. 

\begin{figure}[tbp]
    \centering
    \includegraphics[width=0.95\columnwidth]{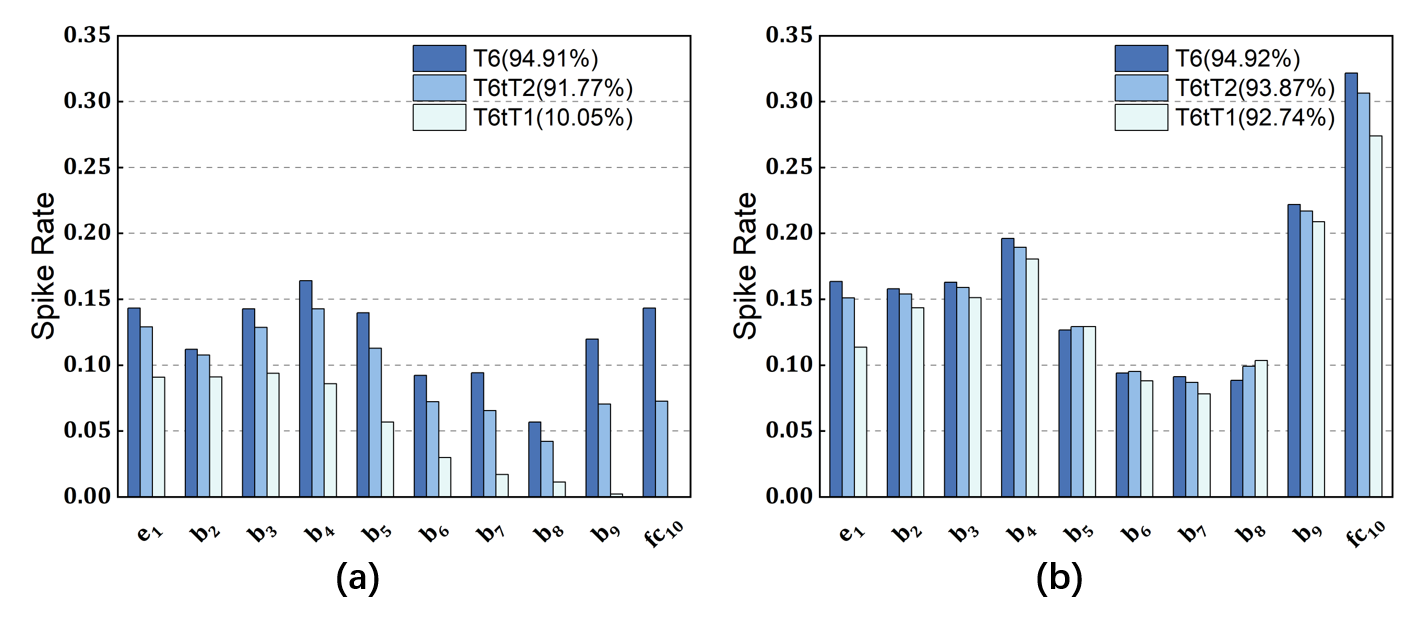}
    \caption{ 
    Impact of inference timesteps on the spike firing rates of neurons in each layer in the (a) SDT and (b) TET methods. Here, $e$ is the encoding layer, $b$ is the middle layer and $fc$ is the fully connected layer.
    }
    \label{fg:spk_rate_sdt-tet}
\end{figure}

\subsubsection{Temporal Compression Based on TET }   
TET applies optimization on each timestep, as \eqref{EQ_TET1}, providing the network with greater temporal robustness and scalability. In contrast, SDT optimizes only at the final timestep, as \eqref{EQ_SDT2}, thereby neglecting the changes in information across different timesteps. Compared to SDT, TET increases the gradient norm near sharp local minima, as \eqref{EQ_TET2}, helping to mitigate the vanishing gradient problem and facilitating the search for a flatter local minimum, which is generally linked to improved generalization performance.
\begin{equation}
\label{EQ_TET1}
\mathcal{L}_{\mathrm{TET}}=\frac{1}{T} \sum_{t=1}^{T} \mathcal{L}_{\mathrm{CE}}[\boldsymbol{O}(t), \boldsymbol{y}] .
\end{equation} 
\begin{equation}
\label{EQ_TET2}
\frac{\partial \mathcal{L}_{\mathrm{TET}}}{\partial W}=\frac{1}{T} \sum_{t=1}^{T}[S(\boldsymbol{O}(t))-\hat{\boldsymbol{y}}] \frac{\partial \boldsymbol{O}(t)}{\partial W}.
\end{equation} 
We leverage this characteristic of TET to the task of temporal compression. Specifically, we apply both SDT and TET to investigate how different inference timesteps affect neuronal output spikes and inference accuracy. As shown in \figref{fg:spk_rate_sdt-tet}, we use the CIFAR10 dataset with the ResNet19 architecture as an example, and the results for other datasets and networks can be found in \apdref{appa}. We first train a baseline model with an initial 6 timesteps, then reduce the timesteps to 2 and 1. In the case of SDT, the spike firing rates of neurons in most layers decrease significantly, making inference with a single timestep unfeasible. However, with TET, we observe that the spike firing rates of neurons remain stable, and the testing accuracy shows a slight decline compared to the baseline model. Subsequently, we fine-tuned the model to further improve testing accuracy by using the weights of the baseline model as initial values and retraining with a timestep of 1. The detailed procedures for the SDT, TET, and fine-tuning processes are outlined in \algref{alg1}, which is provided in \apdref{appb}.

\subsection{Experimental Results }     
In this section, we validate the effectiveness of our approach by comparing it with other SOTA methods for image classification tasks on the CIFAR10/100~\cite{CIFAR_cifar} and Tiny ImageNet~\cite{TINY} datasets. Following previous studies, we utilize spiking VGG16~\cite{VGG}, ResNet19~\cite{ResNet}, and ResNet34 architectures. 
% Detailed settings for hyperparameters and data augmentation methods can be found in Appendix B. 
\tabref{tab:al_t2} presents the comparative results of various temporal pruning methods for SNNs. These methods include the hybrid training approach from~\cite{b4,b5}, the STBP-tdbn method from~\cite{arf2}, the dynamic timesteps method from~\cite{b6}, the conversion technique from~\cite{arf5,arf7}, TEBN from~\cite{arf6}, and the SNN2ANN approach from~\cite{arf8}. As shown in \tabref{tab:al_t2}, our performance at a single inference timestep is comparable to that of other SOTA methods. Notably, the accuracy of ~\cite{b6} is slightly higher than that of our approach on CIFAR10 (CIFAR100), with a lead of 0.31\% (0.44\%). However, its average inference timesteps are slightly greater than ours, at 0.08 (0.12). Although the approach in ~\cite{b6} effectively reduces the number of memory accesses for membrane potentials, it does not eliminate the associated storage and access operations. Similarly, the approach in ~\cite{arf6} shows marginally higher accuracy than our method, exceeding it by 0.08\%, but it requires two inference timesteps. These results indicate that our method achieves a favorable balance of high accuracy and low latency. 
% outperforming other approaches.  

\begin{table}[ht]
\renewcommand{\arraystretch}{1.25}
\centering\caption{\label{tab:al_t2}Comparison with SOTA Temporal Pruning Works.}
\begin{tabularx}{\columnwidth}{XCCCCC} 
\toprule
 & Dataset & Training Method & Model & Accuracy & Timesteps \\
\midrule
~\cite{b4} & CIFAR10& Hybrid & VGG16 & 92.70 & 5\\
~\cite{arf2} & CIFAR10& STBP-tdbn & ResNet19 & 92.34 & 2\\
~\cite{b5} & CIFAR10& IIR-SNN & ResNet20 & 91.10 & 1\\
~\cite{b6} & CIFAR10& SEENN-I & VGG16 & 94.07 & 1.08\\
~\cite{arf5} & CIFAR10& STBP-tdbn & VGG16 & 86.53 & 1\\
Ours & CIFAR10& STI-SNN & VGG16 & 93.76 & 1\\
Ours & CIFAR10& STI-SNN & ResNet19 & 93.74 & 1\\
~\cite{b4} & CIFAR100& Hybrid & VGG16 & 69.67 & 5\\ 
~\cite{b5} & CIFAR100& IIR-SNN & VGG16 & 70.15 & 1\\
~\cite{b5} & CIFAR100& IIR-SNN & ResNet20 & 63.30 & 1\\
~\cite{b6} & CIFAR100& SEENN-I & VGG16 & 71.87 & 1.18\\
~\cite{arf5} & CIFAR100& ANN-SNN & VGG16 & 61.41 & 1\\
~\cite{arf6} & CIFAR100& TEBN & ResNet19 & 75.86 & 1\\ 
Ours & CIFAR100& STI-SNN & VGG16 & 71.43 & 1\\
Ours & CIFAR100& STI-SNN & ResNet19 & 75.78 & 1\\
~\cite{arf7} & Tiny ImageNet & Hybrid & VGG16 & 51.92 & 150\\
~\cite{arf8} & Tiny ImageNet & S2A-STSU & ResNet17 & 56.91 & 15\\ 
~\cite{arf8} & Tiny ImageNet & S2A-STSU & VGG13 & 54.91 & 3\\
Ours & Tiny ImageNet & STI-SNN & ResNet19 & 60.75 & 1\\
\bottomrule
\end{tabularx}\end{table}  

\section{Hardware Implementation} \label{hardware}
In this section, we present the overall hardware architecture and outline the OS dataflow for the convolutional layer. To enhance energy efficiency, we introduce a compressed and sorted spike representation, along with the line buffer. To promote flexibility, we illustrate how adjusting the OS dataflow within the PEs enables STI-SNN to support three distinct convolution modes. Finally, to achieve low latency, we introduce a layer-wise pipelining architecture for multiple convolutional layers, coupled with strategies to optimize output channel parallelism at each layer.

\begin{figure*}[tbp]
    \centering
    \includegraphics[width=0.95\linewidth]{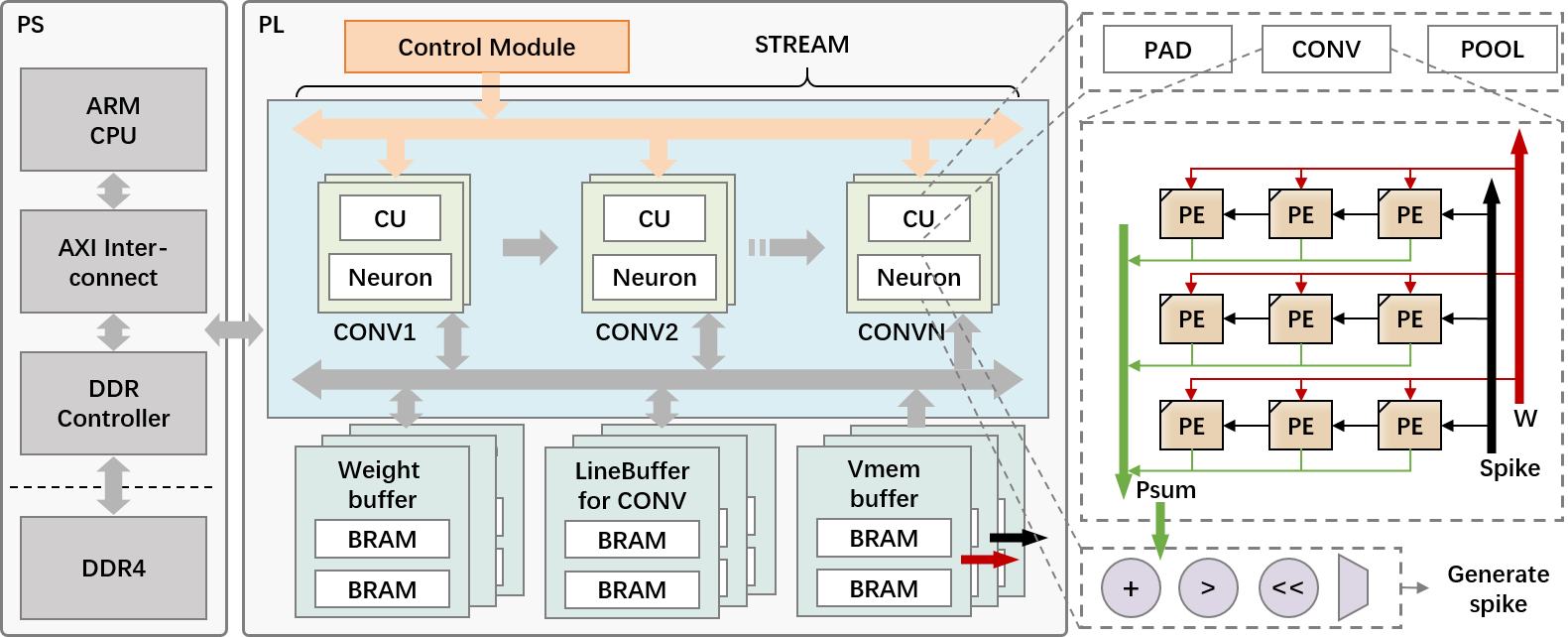}
    \caption{
    The overall architecture of the STI-SNN.
    }
    \label{fg:sti-snn}
\end{figure*}

\subsection{Overall Stream Architecture}
The overall SNN stream architecture is depicted in \figref{fg:sti-snn}. This design incorporates dedicated hardware modules in each layer, optimized for specific computational tasks, thereby enhancing hardware efficiency. Additionally, coarse-grained pipelining across different layers improves parallelism. The convolutional layer serves as the primary computational module, consisting of a computation unit (CU) and a neuron module (Neuron). The CU executes convolution operations on input spikes and weights using a systolic array (SA) formed by multiple PEs. We quantize the weights to 8-bit integers and store them in the on-chip weight buffer, significantly reducing the on-chip storage requirements and memory access costs compared to floating-point format. 
The neuron updates the membrane potential based on the CU's outputs and compares this potential to a threshold to determine whether to fire a spike. The generated spike sequence can be read and processed by the line buffer of the subsequent layer. If the timestep exceeds one, the updated membrane potentials of the neurons must be stored in the Vmem buffer for retrieval in the next timestep.

\begin{figure*}[htbp]
    \centering
    \includegraphics[width=0.95\linewidth]{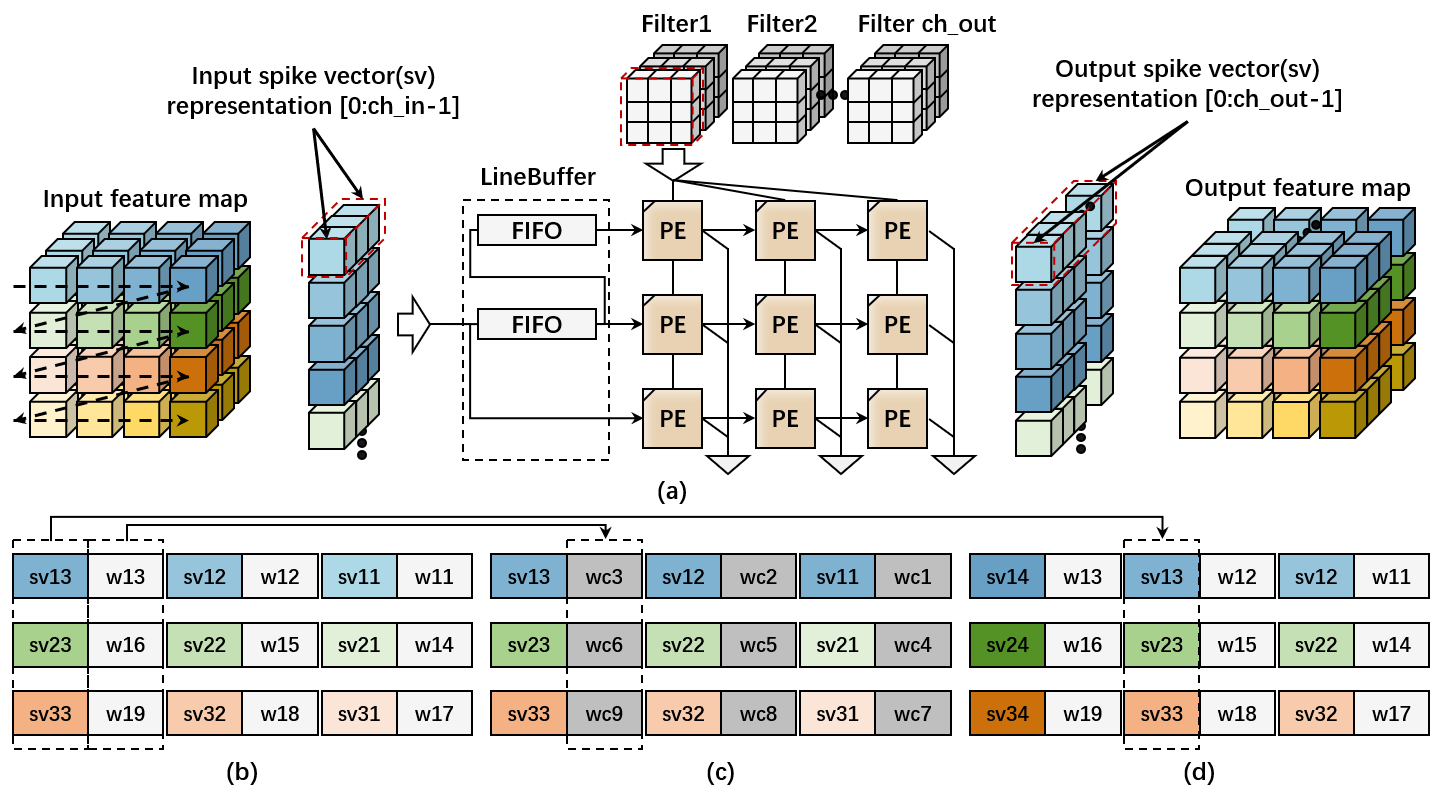}
    \caption{
    The OS dataflow of convolution layers in the STI-SNN. 
    }
    \label{fg:os}
\end{figure*}
  
\subsection{OS Dataflow in the Convolutional Layer of STI-SNN}
The compute array composed of PEs is configured based on the height and width of the filters in each convolutional layer. PEs within the same row can transmit spike vectors to one another. \figref{fg:os} depicts the standard convolution process, in which the input feature map is convolved with each filter, requiring accumulation operations among input channels. As shown in \figref{fg:os}(b), when the input spike vectors are sent to the rightmost PE, the compute array processes spike vectors in the first receptive field in parallel. The weights for each channel of the filter are then accessed sequentially and broadcasted to PEs. The PEs accumulate these weights into their internal membrane potential registers based on the input spikes, as illustrated in \figref{fg:os}(c), where $w_{ck}$ represents the $k\mbox{-}th$ weight in the $c\mbox{-}th$ channel ($k=1 \sim K_w \times K_h$, $c=1 \sim C_i \times C_o$). Once all weights for the $C_o\mbox{-}th$ filter have been processed, an output spike vector is generated and the membrane potential in the registers can be cleared. The compute array then updates the receptive field, as shown in \figref{fg:os}(d). Throughout this process, the input spike vectors require only a single off-chip memory access, and the membrane potential does not need to access higher-level memory. 

\begin{figure}[tbp]
    \centering
    \includegraphics[width=0.95\columnwidth]{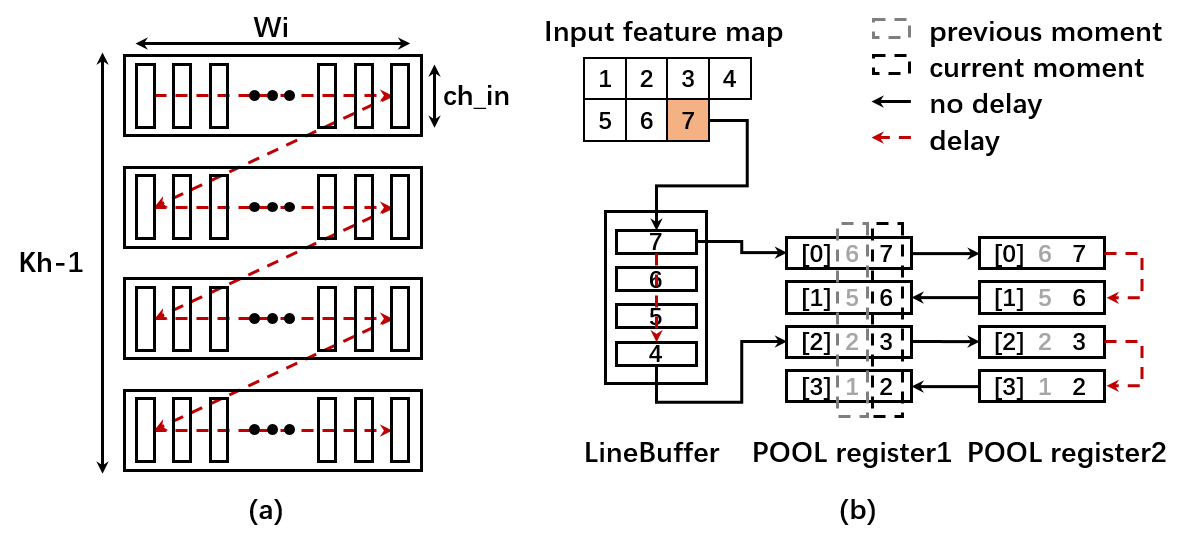}
    \caption{
    (a) Line buffer, (b) pooling layer module.
    }
    \label{fg:linebuffer}
\end{figure}

\begin{table}[tbp]
\renewcommand{\arraystretch}{1.25}
\centering\caption{\label{tab:al_t3}Memory Access Counts for Input Spikes, Weights, and Partial Sums in OS Dataflow Across Various Convolution Modes.}
\begin{tabularx}{\columnwidth}{CCCC} 
\toprule
& Standard Convolution & Depthwise Convolution & Pointwise Convolution \\
\midrule
Input spikes & $H_i \times W_i \times T$ & $H_i \times W_i \times T$ & $H_i \times W_i \times T$\\
Weights & $C_i \times C_o \times H_o \times W_o \times T$ & $C_o \times H_o \times W_o \times T$ & $C_i \times C_o \times H_o \times W_o \times T$\\
Partial Sums & $C_o \times H_o \times W_o \times (T-1)$ & $C_o \times H_o \times W_o \times (T-1)$ & $C_o \times H_o \times W_o \times (T-1)$\\ 
\bottomrule
\end{tabularx}\end{table}  

% \subsection{Hardware Design Considerations} 
\subsection{Compressed and Sorted Spike Representation} 
As shown in \tabref{tab:al_t1}, the number of memory accesses for the input feature map increases linearly with the number of channels, leading to reduced bandwidth utilization. To address this issue, we propose a compressed and sorted spike representation for channels, as illustrated by the input or output spike vectors in \figref{fg:os}. Each spike vector contains spikes from all channels at the same pixel location, organized in channel order. This representation allows for memory access and data transfer using a single spike vector, optimizing efficiency. To further improve the on-chip reuse efficiency, we cache a portion of the input spike vectors required by the PEs in the line buffer. As shown in \figref{fg:os} and \figref{fg:linebuffer}(a), the FIFOs in the line buffer are configured in a tail-to-head arrangement, enabling the tail of one FIFO to connect to the head of the next, and each row of FIFOs simultaneously transmits spike vectors to the corresponding row of PEs. For standard convolution, the number of FIFOs required is $K_h$ where $K_h$ is the height of the convolution kernel. Each FIFO must have a depth of at least $W_i$ and a width of $C_i$ bits, which corresponds to the length of a single spike vector. 
Comparing \tabref{tab:al_t1} and \tabref{tab:al_t3}, we find that off-chip memory accesses for input spikes in OS  dataflow are approximately reduced by $C_i \times K_w \times K_h \times C_o$ times. 

We implement the pooling module based on the line buffer, as shown in \figref{fg:linebuffer}(b). In this setup, rows with indices 0 and 2 in register1 read spike vectors from the start and end of the line buffer, respectively, and transfer this data to register2. After a delay of one clock cycle, rows with indices 1 and 3 in register2 send the shifted data back to register1. All rows in register1 then perform a pooling operation using a logical OR. To ensure the correctness of the operands in the pooling operation, the depth of the line buffer must match the width $W_i$ of the input feature map.

\begin{figure*}[tbp]
    \centering
    \includegraphics[width=0.95\linewidth]{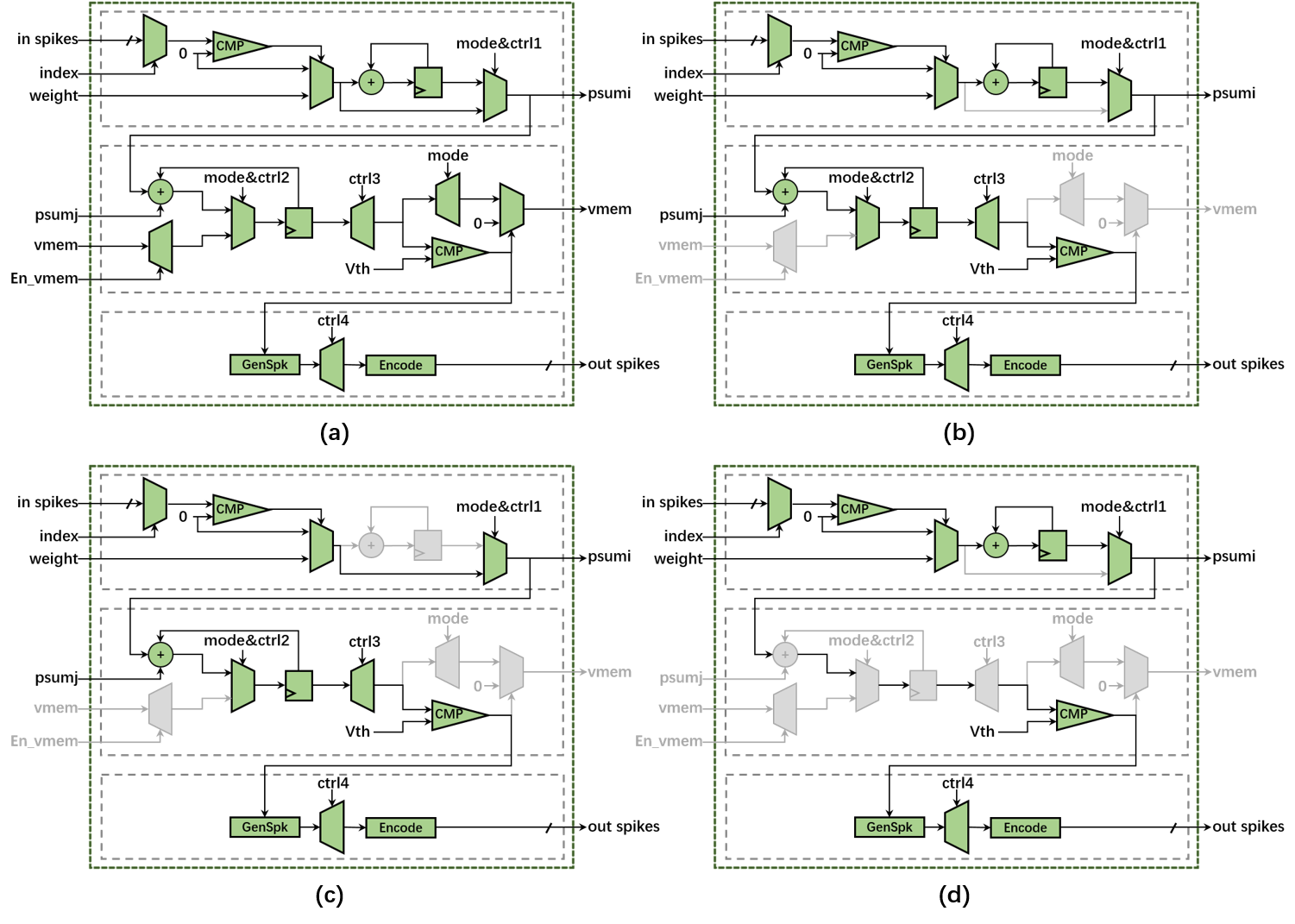}
    \caption{
    PE with multiple convolution modes: (a) PE structure, (b) standard convolution mode, (c) depthwise convolution mode, (d) pointwise convolution mode.
    }
    \label{fg:4pes}
\end{figure*} 

\subsection{Multi-Mode Processing Element} 
In the convolution module, different convolution modes, such as standard convolution, depthwise convolution, and pointwise convolution, affect the OS dataflow, computational efficiency, and memory access efficiency in distinct ways. Depthwise convolution, for instance, aligns the number of channels in the filters with those of the input and output feature maps, significantly reducing the number of parameters and computational complexity compared to standard convolution. As shown in \tabref{tab:al_t3}, the number of on-chip memory accesses for a single layer of weights is reduced by a factor of $C_i$, and the PE does not need to perform cross-channel accumulation operations. Furthermore, a membrane potential register is not required for storage during single timestep inference. Pointwise convolution, utilizing $1 \times 1$ filters, enables adjustment of the number of output channels while reducing the resource overhead of the compute array. To support various convolution modes, it is essential to flexibly adjust the dataflow of the PEs, the size of the compute array, and the processing logic.

\figref{fg:4pes}(a) shows the computing engine designed for convolution operations, comprising three vertically arranged modules: spike accumulation, generation, and output feedforward. This system supports various functionalities based on the configured modes, including single-timestep and multi-timesteps modes, as well as traditional convolution, depthwise convolution, and pointwise convolution modes. In multi-timesteps mode, it loads the historical membrane potential into the registers and saves the updated membrane potential after accumulation. \figref{fg:4pes}(b) shows the traditional convolution mode, where the PEs in the spike accumulation module accumulate weights only upon receiving a spike. Once all spikes from the input channels are processed, the control signal ctrl1 is asserted, allowing each PE to output its accumulated result, referred to as psum. Subsequently, the spike generation module aggregates all psums, and the control signal ctrl3 is asserted to update the output spike sequence based on threshold comparisons. After traversing all output channels, the control signal ctrl4 is asserted to encode the generated spike sequence, yielding the final output. \figref{fg:4pes}(c) shows the depth convolution mode of the PEs, in which they do not accumulate weights on the input channels but directly output the loaded weights upon receiving a spike. \figref{fg:4pes}(d) presents the pointwise convolution mode, which differs from traditional convolution in that the spike generation module does not accumulate psums from the PEs but instead directly compares these results to a threshold and updates the output spike sequence accordingly.

\begin{figure}[tbp]
    \centering
    \includegraphics[width=0.95\linewidth]{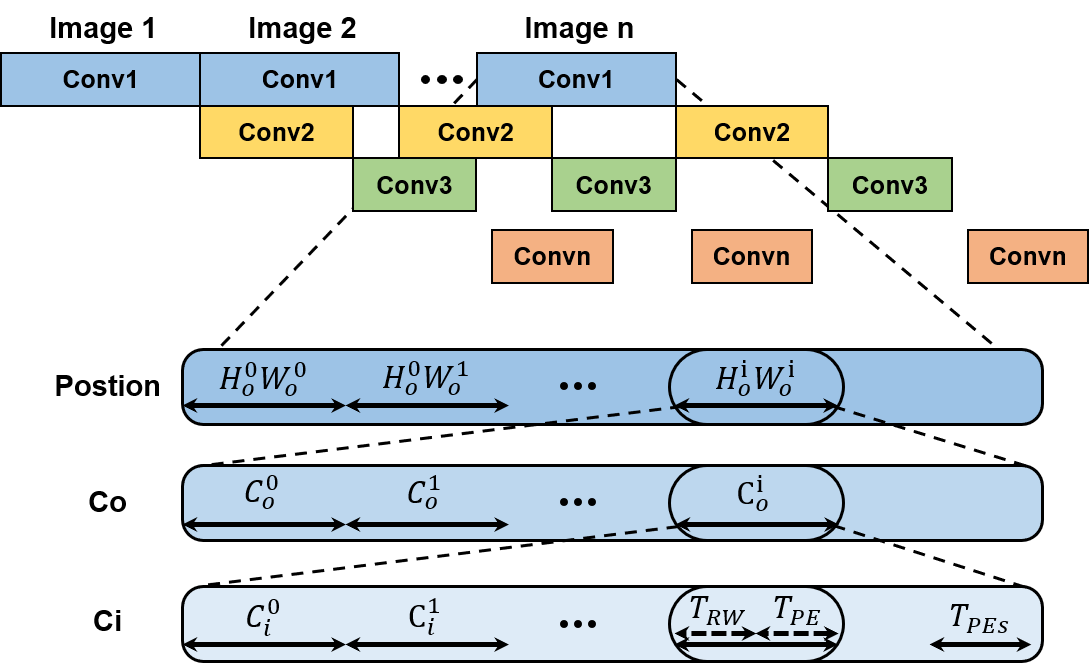}
    \caption{
    Layer-wise pipelining and convolution layer latency analysis.
    }
    \label{fg:pipl}
\end{figure}

\subsection{Inter-layer and Intra-layer Parallelism} 
Thanks to the lightweight algorithm design and FPGA reconfigurability, STI-SNN is implemented as a streaming architecture. In this setup, each computational component of the SNN is assigned a dedicated hardware unit, enabling individual optimization and inter-layer parallelism. We first introduce FIFO buffers to reduce off-chip memory accesses between layers. Additionally, by exploiting the sparse nature of SNN spikes, we encode spike vectors into events, reducing the on-chip memory load for intermediate results. Finally, we optimize output channel parallelism at each layer to improve overall performance and hardware efficiency.

\subsubsection{Layer-wise Pipelined Architecture Design} 
The STI-SNN design uses a layer-wise pipelined architecture, with inter-layer parallelism enhancing system performance, as shown in \figref{fg:pipl}. Assuming that the maximum latency $T_{c_i}$ is in the $i\mbox{-}th$ convolution layer (index $c_i$ is different from $C_i$) and the number of frames processed is $N$, we first compute the total latency $T_{total}$ for the entire pipeline \eqref{EQ_Ttotal}, followed by the calculation of the average latency $T_{avg}$ per frame \eqref{EQ_Tavg}.
\begin{equation}
\label{EQ_Ttotal}
T_{total}=\frac{N \times T_{c_i}+\sum_{j \neq i} T_{c_j}}{N}.
\end{equation} 
\begin{equation}
\label{EQ_Tavg}
 T_{avg}=T_{c_i}+\frac{\sum_{j \neq i} T_{c_j}}{N}.
\end{equation}
Due to the differences in computational complexity and processing speed among layers, we implement a "request-response" mechanism to synchronize inter-layer operations, preventing data loss and processing errors. At the same time, we finely design FIFO buffers for different layers to avoid hardware resource wastage. To optimize data transmission efficiency, we leverage the sparsity of SNN spikes to encode spike events. Combined with a hardware decoder, we only transmit spike events, reducing data volume and improving transmission efficiency. In cases of highly sparse spike signals, this encoding approach is more efficient than transmitting continuous values for each neuron in ANNs. The specific encoding method is: $log_{2}(H_i) + log_{2}(W_i) + C_i$. By integrating the handshake mechanism, FIFO buffers, and spike event encoding, the layer-wise pipelined architecture enhances data transmission efficiency and optimizes hardware resource utilization, balancing performance and power consumption, thereby providing strong support for the application of STI-SNN in energy-efficient computing.   

\subsubsection{Optimization of Output Channel Parallelism} 
In the streaming architecture, the latency of each layer is critical, as the overall processing speed of the pipeline is determined by the convolution layer with the highest latency, as shown in \eqref{EQ_Tavg} and \figref{fg:pipl}. 
Obviously, as $N$ increases, the average latency approaches $T_{c_i}$. To further improve the system's processing efficiency, it is essential to optimize the latency of the convolution layer.
In our proposed design, $T_{c_i}$ can be further decomposed and approximated as shown in \eqref{EQ_Tconv}.
\begin{equation}
\label{EQ_Tconv}
T_{c_i}=H_{o} \times W_{o} \times C_o \times [C_i \times (T_{rw}+T_{pe})+T_{pes}].
\end{equation} 
Here, $T_{rw}$ is the time required to read the weights, $T_{pe}$ is the weight accumulation time within a PE along the input channel direction, and $T_{pes}$ is the accumulation time for the partial sums generated by all PEs. By hiding the weight access time, $T_{rw}$ is eliminated, and $T_{pe}$ is reduced using an addition tree. Additionally, we adopt an output channel parallelization technique to further enhance computation speed. The parallel factors can be independently configured for different convolution layers to achieve a balance between hardware resources and computational efficiency.

\begin{figure}[tbp]
    \centering
    \includegraphics[width=0.95\columnwidth]{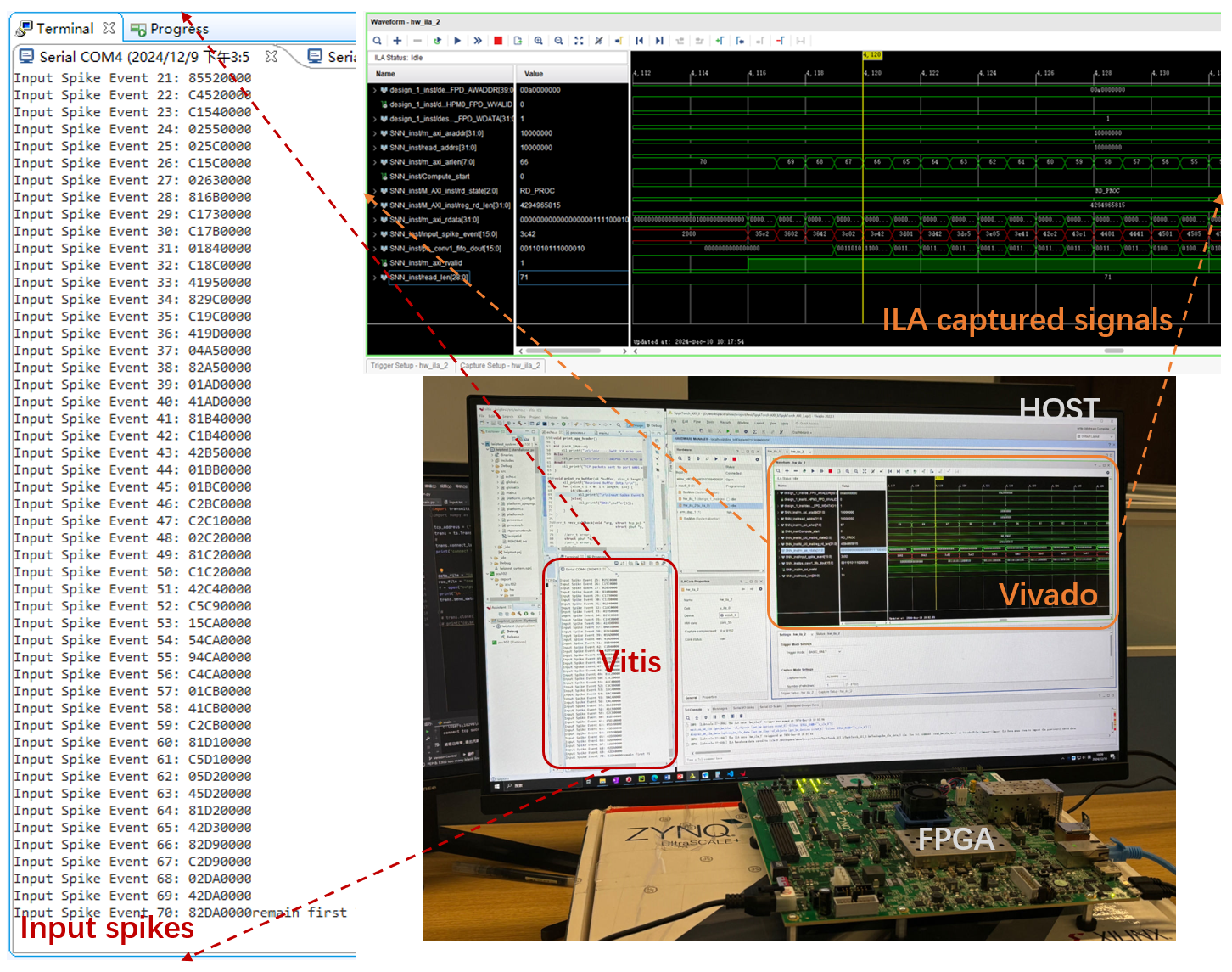}
    \caption{
     Real-world test setup of a classification system for functional validation of the proposed STI-SNN.
    }
    \label{fg:hostandfpga}
\end{figure}  

\section{Implementation and Experimental Results}  \label{results}
We developed corresponding accelerators for three different SNNs on an FPGA platform to evaluate our proposal. We then analyzed their performance and efficiency, comparing them with other FPGA-based SNN accelerators.  

\begin{figure}[tbp]
    \centering
    \includegraphics[width=0.95\columnwidth]{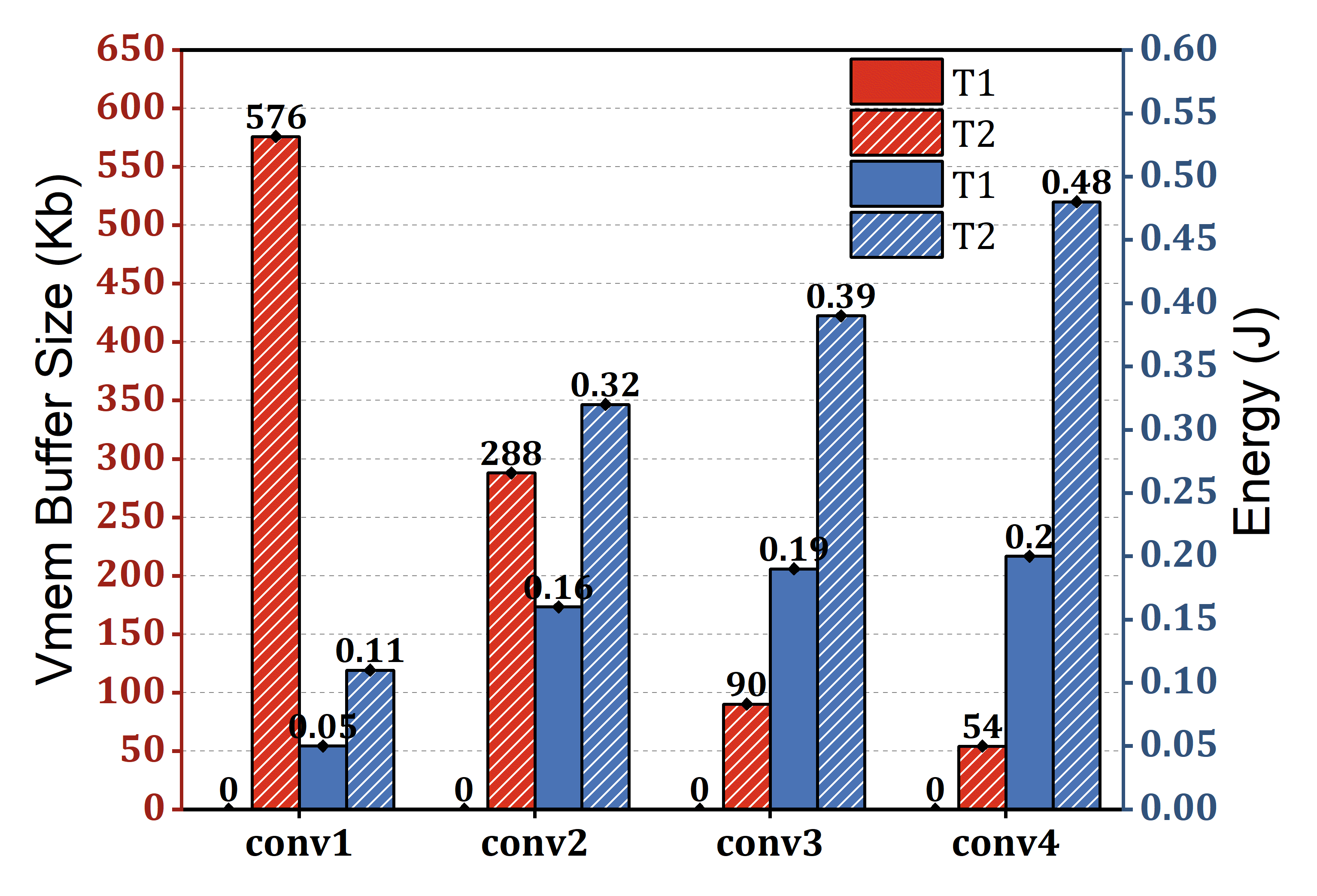}
    \caption{
     Membrane potential memory and energy consumption overhead of each convolution layer in SCNN5 at T1 and T2.
    }
    \label{fg:T1andT2_compare}
\end{figure}  
\begin{figure*}[tbp]
    \centering
    \includegraphics[width=0.95\linewidth]{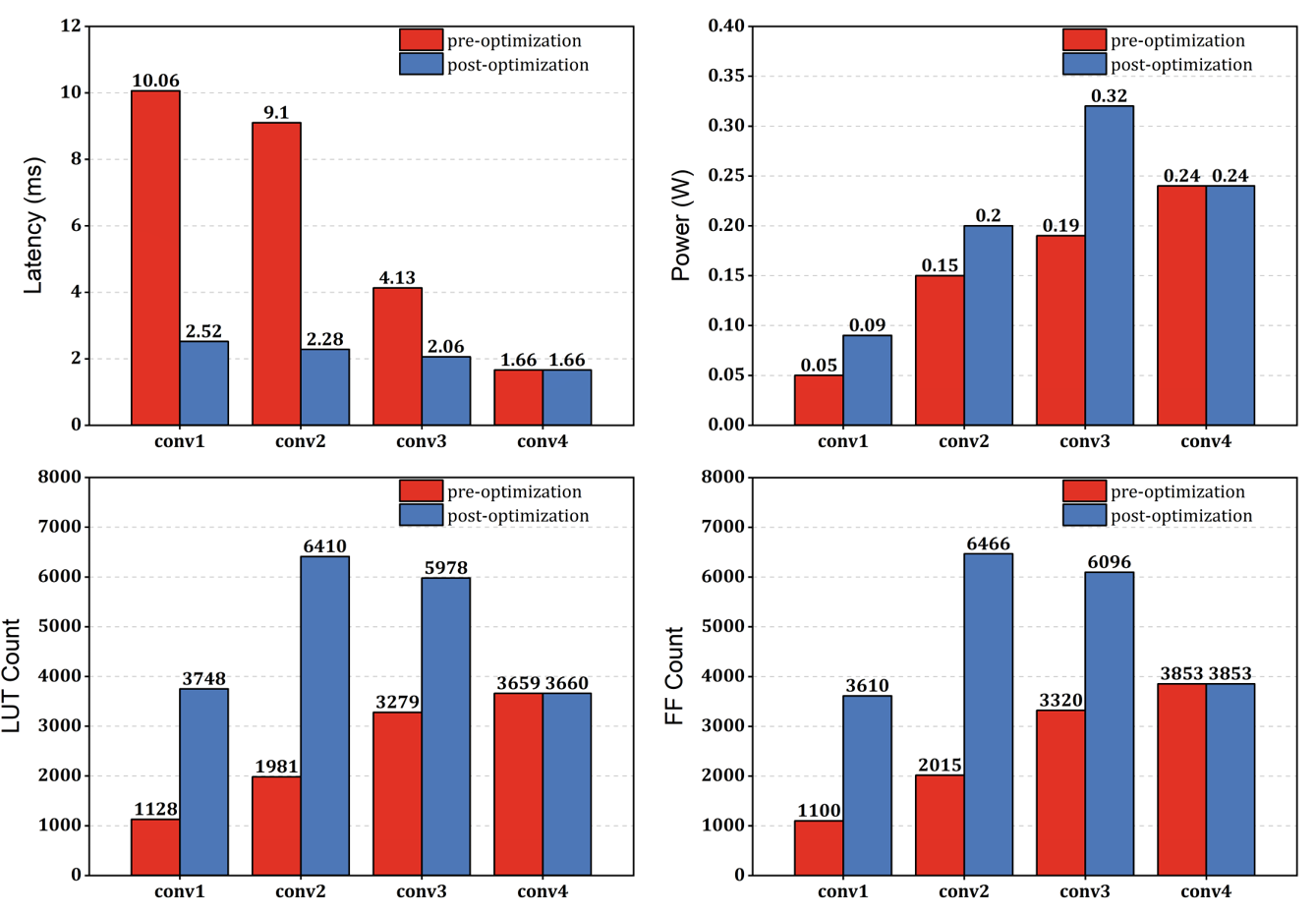}
    \caption{
     Comparison of delay, power, LUT, and FF logic resources before and after output channel parallel optimization in SCNN5.
    }
    \label{fg:parallel_cpmpare}
\end{figure*}

\subsection{Experimental Setup} 
We utilized three SNN models on the MNIST and CIFAR10 datasets for image classification task. For the MNIST dataset, we evaluated a model with the architecture of 28$\times$28 16c3-32c3-p2-32c3-p2-fc (SCNN3) and a variant of MobileNet with the architecture of 28$\times$28 16c3-16dwc3/32c1-32dwc3/64c1-64dwc3/64c1-64dwc3/128c1-fc (vMobileNet). The latter includes a standard convolution layer as the first layer, followed by 4 DSCs and a fully connected layer. For the CIFAR10 dataset, we selected the model with the architecture of 32$\times$32 64c3-p2-128c3-p2-256c3-p2-256c3-p2-512c3-p2-fc (SCNN5). Here, the first convolution layer is used for spike encoding, with the encoded spikes serving as the input to the accelerator. Our testing setup comprises an FPGA board and a host system, as shown in \figref{fg:hostandfpga}. We employed the Xilinx ZCU102 development board as the target FPGA platform, which features an xczu9eg chip with 4.5 GB of DDR4 DRAM. The host system is equipped with two Xilinx tools: Vivado 2021.1 and Vitis 2021.1. Vivado is utilized to develop the corresponding Verilog code for the hardware architecture, as well as simulate, synthesize, and generate the bitstream. Vitis, on the other hand, is used to develop a TCP server that receives input data and control signals from the host and sends results back to it. The host system is responsible for importing the CNN model and generating the model parameters and input images.

\subsection{Overall Performance Analysis} 
\subsubsection{Effects of Inference Timesteps on Energy Consumption and Storage Resources} 
\figref{fg:T1andT2_compare} presents the results of SCNN5 for two inference timesteps, T1 and T2. At T2, membrane potential memory usage decreases from the first to the fourth convolution layer, whereas energy consumption gradually increases. This is due to earlier layers generating more membrane potentials, while later layers have more weights. In contrast, T1 eliminates the need for on-chip storage of membrane potentials across all convolution layers, saving 126 KB in storage resources. Deeper networks can achieve even greater reductions. Additionally, the total energy consumed by the four convolution layers is 0.6J at T1 and 1.3J at T2, indicating that energy consumption decreases linearly with the timesteps for the same number of processed samples.

\subsubsection{Effects of Inter-layer and Intra-layer Parallelism Optimization on Performance and Resource Utilization} 
For SCNN5, implementing a layer-wise pipelining architecture significantly reduced the image inference delay from 24.95ms to 10.06ms. Further optimization of output channel parallelism, with parallel factors set to (4, 4, 2, 1) for the convolution layers, decreased the delay to 2.52ms, achieving a 9.9$\times$ improvement, as shown in \figref{fg:parallel_cpmpare}. We also compared resource utilization and power consumption before and after STI-SNN parallelization. The results indicated that for convolution layers conv1 to conv3, both LUT and FF logic resources, as well as power consumption, increased with the parallel settings. In contrast, conv4, which did not utilize parallel processing, exhibited no changes in resource utilization or power consumption.

\subsection{Comparison and Analysis with Previous Work}
We utilized the Vivado Design Suite for the synthesis, placement, and routing of the accelerator, yielding results on resources, timing, and power. \tabref{tab:al_t4} and \tabref{tab:al_t5} compare the performance of STI-SNN with other SOTA accelerators, focusing on metrics such as accuracy, throughput, accelerator's efficiency, and hardware resource utilization across the same benchmark datasets (MNIST, CIFAR10). For SCNN3, the parallel factor is set to (4, 2) with a total of 54 PEs, while SCNN5 has a parallel factor of (4, 4, 2, 1) and 99 PEs. Notably, vMobileNet is not parallelized. As shown in \tabref{tab:al_t4}, compared to the case without applying output channel parallelism optimization, the inference speeds of the SCNN3 and SCNN5 accelerators are improved by 3.91$\times$ and 4.0$\times$, respectively, with efficiency improvements of 3.64$\times$ and 3.49$\times$. Additionally, the proposed accelerator showcases exceptional flexibility, adapting to various model sizes and network architectures while supporting independent parallel factors for different convolution layers, thus enhancing PE utilization. The efficiency per PE for SCNN3 and SCNN5 is 0.19 and 0.14 GOPS/W/PE, respectively, indicating optimal performance. Resource utilization metrics for the ZCU102 FPGA and comparisons with other works are presented in \tabref{tab:al_t5}.

\begin{table*}[htbp]
    \renewcommand{\arraystretch}{1.25}
    \begin{threeparttable}[b]
        \caption{\label{tab:al_t4}Comparison of STI-SNN with Other SOTA Accelerators on Accuracy, Throughput, Power, and Efficiency Metrics.}
        % \begin{tabularx}{\linewidth}{CCCCCCCCC} 
        \begin{tabularx}{\linewidth}{ccccccccc} 
            \toprule
            &   \begin{tabular}[c]{@{}c@{}} Network \\structure \end{tabular} &   Datasets & \begin{tabular}[c]{@{}c@{}}Accuracy (\%)\end{tabular}  & FPS &  \begin{tabular}[c]{@{}c@{}}kFPS$\times$MOPs \\(GOPS)\end{tabular} & \begin{tabular}[c]{@{}c@{}}Power \\(W)\end{tabular} & \begin{tabular}[c]{@{}c@{}}Efficiency \\(GOPS/W)\end{tabular} & \begin{tabular}[c]{@{}c@{}}Efficiency/PE \\(GOPS/W/PE) \end{tabular} \\
            \midrule
             Fang et al.~\cite{fang} & \begin{tabular}[c]{@{}c@{}} 28$\times$28-32c3-p2-32c3-p2-\\256-10 \end{tabular}&MNIST&99.2&133&0.65&4.5&0.14&-\\
             Ye et al.~\cite{ye} & \begin{tabular}[c]{@{}c@{}} 32$\times$32-32c3-p2-32c3-p2-\\256-10 \end{tabular} &SVHN&82.15&826.4&5.26&0.98&5.35&0.02\\
             Ju et al.~\cite{ju} & \begin{tabular}[c]{@{}c@{}} 28$\times$28-64c5-p2-64c5-p2-\\128-10 \end{tabular} &MNIST&98.94&164&2.50&4.6&0.54&-\\
             Cerebron~\cite{cerebron} & ConvNet\tnote{1}&MNIST&99.40&38500&40.1&1.4&28.6&0.11\\
             Cerebron~\cite{cerebron} & ConvNet&CIFAR10&91.90&94&44.2&1.4&31.57&0.12\\
             Firefly~\cite{firefly} & SCNN-5\tnote{2}&MNIST&98.12&2036&265.76&2.55&104.22&0.05\\
             Firefly~\cite{firefly} & SCNN-7\tnote{3}&CIFAR10&91.36&966&274.49&2.55&107.64&0.05\\
             \textbf{Ours-1}\tnote{4} & SCNN3 &MNIST&99.51&341.3&1.85&0.66&2.79&0.16\\
             \textbf{Ours-2}\tnote{5} & SCNN3&MNIST&99.51&1333&7.22&0.71&10.15&0.19\\
             \textbf{Ours-3} &  SCNN5&CIFAR10&90.31&99.4&5.16&1.34&3.86&0.11\\
             \textbf{Ours-4}\tnote{6} & SCNN5&CIFAR10&90.31&397&20.6&1.53&13.46&0.14\\
             \textbf{Ours-5} & vMobileNet&MNIST&98.11&290&0.75&0.74&1.01&0.03\\
            \bottomrule
        \end{tabularx}
        
        \label{tab:gops}
        \begin{tablenotes}
            \item[1] Classification network with 28$\times$28-16c3-32c3-16c3-10 structure.
            \item[2] Classification network with 28$\times$28-16c3-64c3-p2-128c3-p2-256c3-256c3-10 structure.
            \item[3] Classification network with 32$\times$32-16c3-64c3-p2-128c3-128c3-p2-256c3-256c3-p2-512c3-10 structure.
            \item[4] Using layer-wise pipelining, but not yet implementing output channel parallelism.
            \item[5] Using layer-wise pipelining and output channel parallelism, with parallel factors of 4 2 for the convolution layers.
            \item[6] With parallel factors of 4 4 2 1 for the convolution layers.
        \end{tablenotes}
        
    \end{threeparttable}
\end{table*}  
% \begin{tabular}[c]{@{}c@{}}\end{tabular}

\begin{table}[htbp]
    \renewcommand{\arraystretch}{1.25}
    \centering\caption{\label{tab:al_t5}Resource Utilization and Comparison.}
    \begin{tabularx}{\columnwidth}{CCCCC} 
    \toprule
        &\begin{tabular}[c]{@{}c@{}}Ye et al.\\ \cite{ye}\end{tabular} &\begin{tabular}[c]{@{}c@{}}Cerebron\\ \cite{cerebron}\end{tabular} & \begin{tabular}[c]{@{}c@{}}FireFly \\ \cite{firefly}\end{tabular} & \textbf{Ours}\\
        \midrule
        Device&xc7k325t&xc7z100&xczu3eg&xczu9eg\\
        Dataflow&OS&OS&WS&OS\\
        Precision&FIX16&-&Int8&Int8\\
        Neuron Type&EPC-LIF&IF&IF/LIF&IF\\
        \begin{tabular}[c]{@{}c@{}}Frequency\\(MHz)\end{tabular}&200&200&300&200\\
        \begin{tabular}[c]{@{}c@{}}PE \\Array Size\end{tabular}&256&256&2304&54/99/40\\
        \begin{tabular}[c]{@{}c@{}}Available \\LUT(K)\end{tabular}&203&227&70&274\\
        \begin{tabular}[c]{@{}c@{}}Used \\LUT(K)\end{tabular}&16&86&15&3.5/25.52/7.7\\
        \begin{tabular}[c]{@{}c@{}}LUT \\Utilizaiton(\%)\end{tabular}&7.88&31.05&21.43&1.26/9.31/2.82\\
        Available BRAM&445&755&216&912\\
        Used BRAM&220&283&162&11.5/527.5/13.5\\
        \begin{tabular}[c]{@{}c@{}}BRAM \\Utilizaiton(\%)\end{tabular}&49.44&37.48&75&1.26/57.84/1.48\\ 
    \bottomrule
    \end{tabularx}
\end{table} 

\section{Conclusion} \label{conclusion}
In this paper, we present the STI-SNN, which utilizes a co-design approach integrating algorithms and hardware to achieve high energy efficiency, flexibility, and low latency in resource-constrained scenarios. Firstly, STI-SNN supports inference within a single timestep. At the algorithm level, we introduce a new temporal pruning approach based on the TET loss function. At the hardware level, the accelerator employs an OS dataflow. The co-design of algorithms and hardware eliminates the need for off-chip storage of membrane potentials and associated memory operations. To further optimize the OS dataflow, we propose a compressed and sorted representation of input spikes, which are then cached in a line buffer, reducing the memory access cost of input data and improving on-chip data reuse efficiency. Secondly, STI-SNN supports various network architectures. By adjusting the computation modes of processing elements and parameterizing the configuration of the computation array, STI-SNN can also accommodate DSC models. Lastly, STI-SNN supports both inter-layer and intra-layer parallel processing of data, further reducing inference latency. Overall, through algorithm and hardware co-design, STI-SNN effectively reduces latency and energy consumption while ensuring inference accuracy, providing new insights and practical solutions for designing efficient and flexible SNN accelerators.

{\appendices
\section{} \label{appa}
\figref{fg:SFR} reflects the impact of SDT and TET based temporal pruning on the neurons' spike firing rates and test accuracy across different datasets and networks. When reducing the inference timesteps, TET shows more stable neurons' spike firing rates and better test accuracy compared to SDT. 
\begin{figure*}[tbp]
    \centering
    \includegraphics[width=0.95\linewidth]{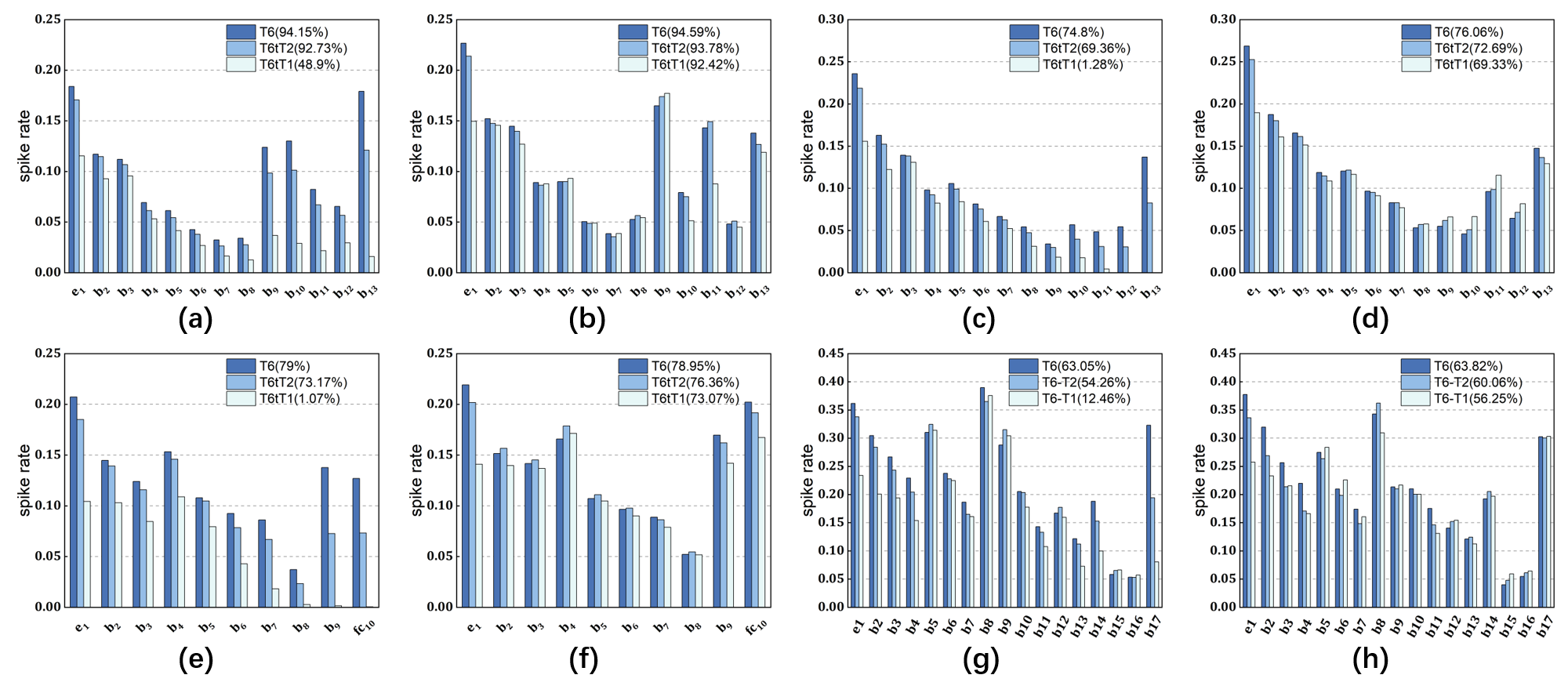}
    \caption{
    The impact of SDT and TET on the neurons' spike firing rates as well as test accuracy across various datasets and networks: (a) CIFAR10 on VGG16 (SDT), (b) CIFAR10 on VGG16 (TET), (c) CIFAR100 on VGG16 (SDT), (d) CIFAR100 on VGG16 (TET), (e) CIFAR100 on ResNet19 (SDT), (f) CIFAR100 on ResNet19 (TET), (g) Tiny Image on ResNet34 (SDT), (h) Tiny Image on ResNet34 (TET).
    }
    \label{fg:SFR}
\end{figure*} 

\section{} \label{appb}
Temporal pruning based on SDT and TET is shown in \algref{alg1}, where, the spike firing rate (SFR) of neurons in each layer of the SNN is calculated by $SFR_l=TotalSpikes_l/N_l$, where $SFR_l$ is the spike firing rate of the neurons in the $l\mbox{-}th$ layer, $TotalSpikes_l$ is the total number of spikes fired by the neurons in the $l\mbox{-}th$ layer over the total timesteps T, and $N_l$ is the total number of neurons in the $l\mbox{-}th$ layer.
\begin{algorithm}[tp]
    \caption{SDT and TET based Temporal Pruning.}
    \label{alg1} 
    \KwIn{initial inference timesteps ($T$), training epochs ($E$), training iterations per epoch ($I_t$)}
    \KwOut{Best model weights ($W_2$)}
    
    Train network with T timesteps \\
    
    \For{$epoch = 1$ \KwTo $E$}{  
        \For{$i = 1$ \KwTo $I_t$}{       
            Get mini-batch training data $X_i$ and class label $Y_i$ \\
            Compute the SNN output $O_i(t)$ of each timestep \\
            
            \If{SDT}{
                Calculate loss function $L = L_{SDT}$ \\
            }
            \If{TET}{
                Calculate loss function $L = L_{TET}$ \\
            }
            
            Backpropagation and update model parameters \\
        }
        Calculate test accuracy and save the best model $W_1$ \\
    }
    
    Based on the pretrained model $W_1$, reduce the inference timesteps directly to $T_{de}$ \\

    \If{SDT}{
        Calculate the spike firing rate of each layer, $SFR_{SDT}$ \\
    }
    \If{TET}{
        Calculate the spike firing rate of each layer, $SFR_{TET}$ \\
    }
    
    Fine-tune the model at timesteps $T_{de}$ \\ 
    Calculate test accuracy and save the best model $W_2$ 
\end{algorithm}
}

\bibliographystyle{IEEEtran} 
{\small
\bibliography{STISNN}  
}

\end{document}